\documentclass[11pt,oneside,final,a4paper]{article}
\usepackage{amsmath,amssymb,makeidx,amsfonts,latexsym,theorem,enumerate}

\newtheorem{lemma}{Lemma}[section]
\newtheorem{theorem}{Theorem}[section]
\newtheorem{corollary}{Corollary}[section]
\newtheorem{proposition}{Proposition}[section]

\DeclareMathOperator{\dist}{dist}

\begin{document}

\author{
Sergio Albeverio \\ Institute of Applied Mathematics \\ Bonn University,
531 55 Bonn, Germany
\and
Matthias Gundlach \\ Institute for Dynamical Systems \\
University of Bremen, P.O. Box 330 440, 28 334 Bremen, Germany  
\and
Andrei Khrennikov and Karl-Olof Lindahl\\
Department of Mathematics, Statistics and Computer
Science \\ V\"{a}xj\"{o} University, 351 95 V\"{a}xj\"{o}, Sweden}

\title{On Markovian behaviour of $p$-adic random dynamical systems}
\date{\today}
\maketitle

\begin{abstract} We study Markovian and non-Markovian behaviour of stochastic processes
generated by $p$-adic random dynamical systems. Given a family of $p$-adic monomial random 
mappings generating a random dynamical system.
Under which conditions do the orbits under such a random dynamical system
form Markov chains? It is necessary that the mappings are Markov
dependent. We show, however, that this is in general not sufficient.
In fact, in many cases we have to require that the mappings are
independent.
Moreover we investigate some geometric and algebraic properties for $p-$adic monomial
mappings as well as for the $p-$adic power function which are essential to the
formation of attractors.
$p$-adic random dynamical systems can be useful in so called $p$-adic quantum phytsics
as well as in some cognitive models.
\end{abstract}

\section{Introduction}

In this paper the state space of a dynamical system will be the field of $p-$adic numbers. The 
$p-$adic numbers are basically the rational numbers together with a ''$p-$%
adic absolute value'' whose properties differ (strongly) from the ones of
the usual
absolute value. The $p-$adic numbers were explicitly first studied by K.
Hensel at the end of the nineteenth century. For a long time they were only
considered as a branch of pure mathematics. However, in the last decade,
there has been an increasing interest for $p-$adic numbers in theoretical
physics and biology \cite{Beltrametti Cassinelli}-\cite{Albeverio Khrennikov Kloeden}.

The theory of dynamical systems is basically concerned with the study of the long-term
behavior of systems. Formally, a system has two components; 1) a \emph{%
state space} $X$ : the collection of all possible states of the system, 2) a 
\emph{map} $\psi :X\rightarrow X$ from $X$ into itself representing the
evolution of the system where the state $x_1=\psi x$ is taken as the state
at time $1$ of a system which at time $0$ was in $x.$ The state $x_0=x$ is
called the \emph{initial state} or the initial condition of the system. In
this way a state $x_n$ is mapped into the state $x_{n+1}=\psi x_n$ where $%
x_n=\psi ^nx$ represents the state of the system at time $n$ which at time $%
0 $ was in state $x.$

Systems like these are deterministic in the sense that given the initial
state $x$ and the map $\psi $ one can foresee the whole future of the system
which can be represented by the \emph{orbit,} $\{x,\psi x,\psi ^2x,...,\psi
^nx,...:n\in \mathbb{Z}^{+}\},$ of $x$ under $\psi $. Such models may work very
well for isolated systems not perturbed by noise. But in general such models
are inadequate. We have to take into account some influence of noise on the
system. Therefore we let the map $\psi $ depend on time, $n,$ and a random
parameter $\omega $ so that $\psi =\psi (n,\omega ).$ We will study models
which involve the concept of a random dynamical system. Roughly speaking, a
random dynamical system is a mechanism which at each time $n$ randomly
selects a mapping $\psi (n,\omega )$ by which a given state $x_n$ is mapped
into $x_{n+1}=\psi (n,\omega )x_n.$ The mappings are selected from a given
family $(\psi _s)_{s\in S}$ of mappings for some index set $S.$ Thus $(\psi
_s)_{s\in S}$ is the set of all realizable mappings. The selection mechanism
is permitted to remember the choice made at time $n,$ $i.e.$ the probability
of selecting the map $\psi _s$ at time step $n+1$ can depend on the choice
made at time $n.$ To model the selection procedure we use another system, a
metric dynamical system, see next section.

For a random dynamical system we
can only predict what will \emph{probably} happen to the system in the
future. Now, suppose that we find the system in the state $x_n$ at time $n.$
What is the probability of observing the state $x_{n+1}$ in the next time
step? The answer to this question may depend on our knowledge of the history
of the system. In this paper we investigate under what condition we do not
need to know anything about its history, except possibly its initial state,
to predict the probability of its future behavior. This investigation is 
based on the work in \cite{Lindahl}. Systems which behave in
this way, $i.e.$ the future behavior is independent of the past and depends
only on the present state, are called Markov processes and are
more easy to handle in scientific research. This is one of the reasons why
physics has developed as it has.

In the long-term behavior of a system two things may happen: $1)$ Almost
every possible state of the system is reached from almost every initial state
(ergodicity). $2)$ The dynamics is attracted to an attractor $A$ of states
in the sense that there is a subset, $U$ of $X,$ properly containing $A$ and
consisting of states which tend to $A$ as time goes to infinity, $i.e.$ $%
\lim_{n\rightarrow \infty }\psi ^nu\in A$ for every $u$ belonging to $U.$ In
the random case an attractor $A$ may depend on the random parameter $\omega $
so that $A=A(\omega ).$

In fact, dynamical systems like those studied in this paper have
been proposed as models for describing some features of the thinking process,
see for example \cite{Khrennikov,Khrennikov et al}. 
In these models the consciousness generates
an idea $x$ (initial state) which evolves in time under a dynamical system
in the subconscious. This system is perturbed by noise, physical and
psychological, in a random manner. The mentioned features of the thinking process
including the noise are then modeled as a $p-$adic random dynamical system. 

In such models $p-$adic integers are used for the coding of cognitive
information. It seems that such a $p-$adic coding describes well the ability
of cognitive systems to form associations. A $p-$adic integer $x=\sum{\alpha _np^n}$
where $\alpha _n\in \{0, 1, \ldots, p-1\}$ (see section \ref{padicnumbers}), is
considered as an information string $x=(\alpha_0, \alpha_1,\ldots , \alpha _n,\ldots )$;
a $p-$adic distance induces the following nearness on the space of such information
strings: $x=(\alpha _j)$ and $y=(\beta _j)$ are close if and only if
$\alpha _0=\beta _0,\ldots, \alpha _N=\beta _N$ and $\alpha _{N+1}\neq \beta _{N+1}$
for a sufficiently large $N$. Thus there is a hierarchical structure between
digits $\alpha _0, \alpha _1,\ldots $ which are used for the coding of an idea
$x=(\alpha _j)$. This structure gives identification of ideas via blocks of
associations $b_0=(\alpha _0)$, or $b_1=(\alpha _0,\alpha _1)$ or
$b_2=(\alpha _0, \alpha _1, \alpha _2)$, $\ldots$.

The Markov property
is a very important characteristic of the process of thinking (or memory
recalling, see \cite{Albeverio Khrennikov Kloeden}). One of the most
interesting consequences of our investigations is that the process of
recalling described by the random dynamical model of \cite
{Khrennikov et al} can be both Markovian or non-Markovian depending on the
choice of the initial idea $x$ (and the prime number $p$).

\subsection{Definition of a random dynamical system}

We will study random dynamical systems in the framework of Arnold, \cite{Arnold}.

\noindent\textbf{Definition (Random dynamical system (RDS))} Let $(X,d)$ be a metric
space with a Borel $\sigma-$algebra. A \emph{measurable
random dynamical system\footnote{%
Random dynamical system(s) are henceforth abbreviated as ''RDS''. }} on the
measurable space $(X,\mathcal{B})$ over a metric DS $(\Omega ,\mathcal{F},%
\mathbb{P},(\theta (t))_{t\in \mathbb{T}})$ with time $\mathbb{T}$ is a mapping 
$\varphi :\mathbb{T}\times \Omega \times X\rightarrow X,\quad (t,\omega ,x)\mapsto
\varphi (t,\omega ,x)$, with the following properties:

\begin{description}
\item  (i) \emph{Measurability}: $\varphi $ is $\mathcal{B}(\mathbb{T})\otimes 
\mathcal{F}\otimes $ $\mathcal{B},$ $\mathcal{B}-$measurable.

\item  (ii) \emph{Cocycle property}: The mappings $\varphi (t,\omega
):=\varphi (t,\omega ,\cdot ):X\rightarrow X$ form a cocycle over $\theta
_{\cdot },$ $i.e.$ they satisfy 
$\varphi (0,\omega )=id_X$ for all $\omega \in \Omega$ if $(0\in \mathbb{T)}$,
and\end{description}
\begin{equation}
\varphi (t+s,\omega )=\varphi (t,\theta (s)\omega )\circ \varphi (s,\omega
)\quad \text{for all\quad }s,t\in \mathbb{T},\quad \omega \in \Omega .
\label{cocycle}
\end{equation}

\subsection{Generation in Discrete Time\label{discrete time and generation}}

Let the random map $\varphi $ be a RDS with one-sided discrete time $\mathbb{T}=%
\mathbb{Z}^{+}.$ Let us introduce the time-one mapping $\psi (\omega ):=\varphi
(1,\omega ).$ The repeated application of the cocycle property forward in
time gives 
\begin{equation}
\varphi (n,\omega )=\left\{ 
\begin{array}{cc}
\psi (\theta ^{n-1}\omega )\circ ...\circ \psi (\omega ), & n\geq 1, \\ 
id_X, & n=0.
\end{array}
\right.   \label{discrete cocycle}
\end{equation}
In this way the metric DS selects a mapping $\psi (\theta ^n\omega ),$ at
each time $n,$ which takes the state $x_n$ to the state $x_{n+1}=\psi
(\theta ^n\omega )x_n.$ Thus we can write the one-sided discrete time
cocycle $\varphi (n,\omega )x$ as the ''solution'' of a random difference
equation 
\begin{equation}
x_{n+1}=\psi (\theta ^n\omega )x_n,\quad n\geq 0,\quad x_0=x\in X.
\label{random difference equation}
\end{equation}
Conversely, given a metric DS $\theta =(\Omega ,\mathcal{F},\mathbb{P},(\theta
(t))_{t\in \mathbb{T}})$ and family of measurable mappings $\psi =(\psi (\omega
))_{\omega \in \Omega }$ from $X$ into itself, such that $(\omega ,x)\mapsto
\psi (\omega )x$ is $\mathcal{F}\otimes \mathcal{B},\mathcal{B-}$
measurable, the map $\varphi $ defined by (\ref{discrete cocycle}) is a
measurable RDS. We say that $\varphi $ is \emph{generated} by $\psi .$

\section{Definition of the monomial RDS}

Monomial RDS are stochastic generalizations of deterministic DS of the form 
\begin{equation}
(X,(\psi _s^n)_{n\in \mathbb{Z}^{+}}),\text{ where }\psi _sx=x^s,\quad s\in 
\mathbb{N},\text{ }x\in X.  \label{deterministic flow}
\end{equation}
In this paper the state space $X\;$ is a subset of the field of $p-$adic
numbers\footnote{%
The field of $p-$adic numbers will be introduced in the next section.}$.$ We
shall introduce perturbations of DS defined by (\ref{deterministic flow}).
This can be done as follows. First, let $s$ depend on chance. That is, we
let $s:\Omega \rightarrow S=\{s_1,...,s_r\}$ be a discrete random variable
defined on a probability space $(\Omega ,\mathcal{F},\mathbb{P})$ equipped with
a measure-preserving and invertible transformation $\theta $. For discrete
time, $\theta $ generates a metric DS, $\theta :=(\Omega ,\mathcal{F},%
\mathbb{P},(\theta ^n)_{n\in \mathbb{Z}})$. Then we let $\theta $ describe
the perturbation of the
random variable $s$ so that $s$ will become a stochastic process. This can be
modeled with a sequence $(S_n),$ of random variables, where 
\[
S_n(\omega )=s(\omega )s(\theta \omega )...s(\theta ^{n-1}\omega ).
\]
The random map $\phi :\mathbb{Z}\times \Omega \times
X\rightarrow X,$ defined by 
\begin{equation}
\phi (n,\omega )x=\left\{ 
\begin{array}{cc}
x^{S_n(\omega )},\qquad  & n\geq 1, \\ 
x,\qquad  & n=0, 
\end{array}
\right.   \label{definition of RDS}
\end{equation}
forms a monomial RDS over the metric DS $\theta .$ Then in the sense of (\ref
{random difference equation}) with $\psi (\theta ^n\omega )x=x^{s(\theta
^n\omega )}$ the cocycle $\phi (n,\omega )x$ can be considered as the
solution of the random difference equation 
\[
x_{n+1}=x_n^{s(\theta ^n\omega )},\quad n\geq 0,\quad x_0=x\in X.
\]

The mappings $\psi (\theta ^n\omega )$ can be generated by a Markov shift
in the following way.
Let $S$ $=\{s_1,...,s_r\}\subset \mathbb{N}$ be the state space of the random
variable $s$ which we now want to define. For this purpose we form the
product space $S^{\mathbb{N}}=\{\omega =(\omega _0,\omega
_1,...):\quad \omega _i\in S\}$ and define the random variable $s$ as the
coordinate map $s:$ $S^{\mathbb{N}}\rightarrow S,\quad $ $\omega \mapsto \omega
_0.$ Then the Markov shift $\theta =(S^{\mathbb{N}},\mathcal{F}(S^{\mathbb{N}}),%
\mathbb{P},(\theta ^n)_{n\in \mathbb{N}})$ over $S^{\mathbb{N}}$ with transition
matrix $P,$ generates a family $(s(\theta ^n\cdot ))_{n\in \mathbb{N}}$ of
random variables (coordinate mappings) by the relation 
\begin{equation}
s(\theta ^n\omega )=\omega _n.  \label{RVs}
\end{equation}
Then we can consider the Markov shift $\theta $ as a mechanism which selects
mappings from the family $\psi=(\psi _{s_1},...,\psi _{s_r})$ of monomial
mappings where $\psi_{s_i}x=x^{s_i}$ and $s_i\in S.$ 
Moreover, by the Markov property of the
Markov shift, the random variables (\ref{RVs}) form a Markov process. In
this way the mappings $(\psi (\theta ^n\omega ))_{n\in \mathbb{Z}^{+}}$ are
Markov dependent. Thus, the role of $S$ is to specify the realizable
mappings. The Markov shift relates their dependence.

\section{State space analysis}

In order to investigate the stochastic properties of a RDS $\phi $ of the
form (\ref{definition of  RDS}) over a Markov shift $\theta$, we first
have to know something about the state space $X$ of $p-$adic numbers and
especially the properties of monomial mappings on $X$. The main consequence
of this section is that the set of roots of unity, $\Gamma _p,$ in $\mathbb{Q}_p$ is
an attractor for RDS $\phi $ and that $\Gamma _p$ is isomorphic to the
multiplicative group in the residue class modulo $p.$

\subsection{$p-$adic numbers\label{padicnumbers}}

By the fundamental theorem of arithmetics every rational number $x\in \mathbb{Q}$
can be written as 
\[
x=p^{ord_p(x)}\frac ab,\quad p\nmid ab, 
\]
for every prime number $p.$ Then every prime number $p$ induces a $p-$adic
valuation $\left| \cdot \right| _p$ on $\mathbb{Q}$; $\left| x\right| _p=p^{-ord_p(x)}$,
with the following properties
$1)$ $\left| x\right| _p=0$ if and only if $x=0$;
$2)$ $\left| xy\right| _p=\left| x\right| _p\left| y\right| _p$ for
every $x,y\in \mathbb{Q}$;
$3)$ $\left| x+y\right| _p\leq \max \{\left| x\right| _p,\left|
y\right| _p\}$ for every $x,y\in \mathbb{Q}$ with equality when\ $\left|
x\right| _p\neq \left| y\right| _p$.
Property $3)$ is stronger than the ''usual'' triangle inequality and is
called the \emph{strong triangle inequality}. For a prime number $p$ the 
$p-$adic valuation induces a metric $d_p$ on $\mathbb{Q}$ defined by $%
d_p(x,y)=\left| x-y\right| _p.$ But the metric space $(\mathbb{Q},d_p)$ is not
complete. The completion of $\mathbb{Q}$ with respect to $d_p$ constitutes the
field of $p-$adic numbers which we denote by $\mathbb{Q}_p$. It turns out that we can
represent $\mathbb{Q}_p$ as the family of all formal sums according to
\begin{equation}
\mathbb{Q}_p=\{x=\sum_{n=N}^\infty a_np^n:a_n\in \{0,...,p-1\},\quad N=N(x)\in 
\mathbb{Z}\}.  \label{qp}
\end{equation}
Let $x$ be a $p-$adic
number with the expansion $x=\sum_{n=N}^\infty a_np^n.$ It can be shown, 
\cite{Gouvea,Schikhof}, that $x$ then has the valuation
$\left| x\right| _p=p^{-k}$,\quad if $a_k\neq 0$ and $a_n=0$ for every $n<k$. 

In other words, the integer $k$ ($\geq N)$ represents the first non-zero
term in the $p-$adic expansion (\ref{qp}) of $x.$ Hence, we need not in
general know every term in the sum (\ref{qp}) to find the valuation of a $p-$%
adic number. We compare this with the valuation on the real numbers, the
absolute value, where we have to know the decimal expansion with infinite
precision.

The $p-$adic integers, which we denote by $\mathbb{Z}_p,$ are $p-$adic numbers
of the form:
$\mathbb{Z}_p=\{x=\sum_{n=0}^\infty a_np^n:a_n\in \{0,...,p-1\}\}. $
Hence the $p-$adic integers coincide with the unit disk, $B_1(0)$. In what
follows we let $S_1(0)$ denote the unit sphere.

It is often useful to consider cosets in $\mathbb{Z}_p.$ Let us form the
multiplicative coset $p\mathbb{Z}_p=\{px:x\in \mathbb{Z}_p\}$.
Then $p\mathbb{Z}_p$ is a maximal ideal (in fact also a
prime ideal) in $\mathbb{Z}_p.$ Let us therefore form the quotient field $\mathbb{Z%
}_p/p\mathbb{Z}_p$ consisting of $p$ additive cosets: 
$p\mathbb{Z}_p,\ 1+p\mathbb{Z}_p,\ ....,\ p-1+p\mathbb{Z}_p$, 
isomorphic to $\mathbb{F}_p;$ the residue class modulo $p$.\\ 
\textbf{Remark 1} There is a correspondence between balls and cosets in $%
\mathbb{Z}_p$ ($\mathbb{Q}_p$) since $i+p\mathbb{Z}_p=\{x\in \mathbb{Z}_p:a_0=i\}=$ $%
B_{1/p}(i).$ Moreover two elements $x$ an $y$ belongs to same coset, $i+p%
\mathbb{Z}_p,$ if and only if $\left| x-y\right| _p\leq 1/p$.

\subsection{Fundamental properties of monomial mappings}

The following lemma reveals some important properties of monomial mappings $%
\psi _s,$ 
\[
\psi _s:\mathbb{Q}_p\rightarrow \mathbb{Q}_p,\quad x\rightarrow x^s\quad \text{,}%
s\in \mathbb{N}, 
\]
on the field of $p-$adic numbers.

\begin{lemma}
\label{lemma 1}
Let $\gamma\in S_1(0)$ and $u\in p\mathbb{Z}_p$.
Then for all natural numbers $n$,
\begin{equation}
\left| (\gamma+y)^n-\gamma^n\right| _p\leq \left| n\right| _p\left| u\right| _p,
\label{fundamental-inequality}
\end{equation}
with equality for $p>2$.
\end{lemma}
{\noindent \bf Proof. }
Let $n=mp^d$, where $p$ does not divide $m$. Define $g\colon x \mapsto x^m$, and
$h \colon x \mapsto x^p$. Then 
\begin{align*}
\left| g(\gamma+u)-g(\gamma)\right| _p &=\left| (\gamma+y)^m-\gamma^m\right|
_p=\left|\sum_{k=0}^m\binom mk\gamma^{m-k}u^k-\gamma^m\right| _p \\
&=\left|mu\gamma^{m-1}+o(u^2)\right| _p = \left|m\right|_p  \left|u\right|_p
 \left|\gamma^{m-1}\right|_p = \left|u\right|_p,
\end{align*}
by the strong triangle inequality (here $o(z)$ means terms of $p-$order smaller
than or equal to the order of $z$, which here is simply all the rest of the
binomial expansion). Thus the map $g$ is an isometry. Set 
$v=g(\gamma+u)-g(\gamma)$, and $y=g(\gamma$). The prime number $p$
divides all the binomial coefficients $\binom pk$ for $1<k<p$, thus we have
for $p>2$
\begin{equation*}
\left| h(y+v)-h(y)\right| _p = \left| pvy^{p-1}+o(pv^2)\right| _p
=\left|p\right| _p\left|v\right| _p=\left|p\right| _p\left|u\right| _p,
\end{equation*}
and for $p=2$ we have
\begin{equation*}
\left| h(y+v)-h(y)\right| _p = \left| pvy^{p-1}+o(v^2)\right| _p
\leq \left|p\right| _p\left|v\right| _p=\left|p\right| _p\left|u\right| _p.
\end{equation*}
Thus, $d$ iterations of $h$ give
\begin{equation*}
\left| (\gamma+y)^n-\gamma^n\right| _p=\left| h^d(g(\gamma+u))-h^d(g(\gamma))\right|
_p \leq \left| n\right| _p\left| u\right| _p,
\end{equation*}
where equality holds for $p>2$. $\Box$

\begin{corollary}
\label{corollary 1}
Let $x,y\in S_1(0)$ and suppose $\left| x-y\right| _p<1.$
Then for all natural numbers $n,$%
\begin{equation}
\left| x^n-y^n\right| _p\leq \left| n\right| _p\left| x-y\right| _p,
\label{fundamental-inequality}
\end{equation}
with equality for $p>2.$
\end{corollary}

{\noindent \bf Proof. }
By hypothesis $x-y\in p\mathbb{Z}_p$. Put $x-y=u$ and $x=\gamma$.
Then the corollary follows directly from Lemma \ref{lemma 1}.
$\Box$\\\textbf{Remark 2} The equality $\left|
\sum\limits_{k=1}^n\binom nky^{n-k}(x-y)^k\right| _p=\left| n\right|
_p\left| x-y\right| _p$ does not always hold for\ $p=2;$ For example $\left|
\sum\limits_{k=1}^4\binom 4k3^{n-k}2^k\right| _2<\left| 4\right| _2\left|
2\right| _2.$ Hence we do not always have equality in (\ref
{fundamental-inequality}) in the case that $p=2$. $\Box $

Let $s$ be a natural number divisible by $p.$ From Corollary \ref{corollary 1} we
see that the corresponding monomial map $\psi _s$ is contracting on
the unit sphere $S_1(0)$ since in this case we have $\left| \psi _sx-\psi
_sy\right| _p\leq 1/p\left| x-y\right| _p.$ We will use a special case ($s=p$%
) to determine all possible fixed points under monomial mappings on $\mathbb{Q}%
_p$.

\subsubsection{Fixed points and roots of unity in $\mathbb{Q}_p$}

In the study of dynamical systems it is important to know the fixed points
of the mappings generating the system. A point $x$ is a \emph{fixed point}
under the monomial map $\psi _s$ if and only if it satisfies the equation $%
\psi _sx=x,$ $i.e.$ if $x^s=x.$ Clearly $0$ is a fixed point under
iterations of $\psi _s$ for all natural numbers $s.$ A fixed point $x\neq 0$
under a monomial map $\psi _s$ satisfies $x^{s-1}=1$ (since $\mathbb{Q}_p$ is a
field every element except $0$ has a multiplicative inverse)$,$ $i.e.$ $x$ is
a root of unity$.$ In $\mathbb{Q}_p$ ( $\mathbb{Z}_p)$ there are $p-1$ roots of
unity. One can show the existence of $p-1$ zeroes to the polynomial $F(x)=$ $%
x^{p-1}-1$ by studying the monomial map $\psi _p:$ $x\mapsto x^p:$

First we observe that each coset in $\mathbb{Z}_p/p\mathbb{Z}_p$ is closed under
iterations of $\psi _p.$ This is a consequence of the fact that
$a^p\equiv a\operatorname{mod}p$. Moreover the condition of Corollary \ref{corollary 1}
is satisfied for every coset$,$ $i+p\mathbb{Z}_p$ where $i\neq 0,$ see Remark 1.
>From this we conclude that $\psi _p$ is a contraction on each of these cosets
(as a consequence of Banach's fixed point theorem).
That $\psi _p$ is a contraction on $p\mathbb{Z}_p=B_{1/p}(0)$ follows from the
strong triangle inequality. Consequently every coset, $i+p%
\mathbb{Z}_p,$ has a unique fixed point, $\xi _i,$ such that $\xi _i^p=\xi _i.$
And if $i$ is different from $0$ we also have that $\xi _i^{p-1}=1.$ This
proves the existence of $p-1$ roots of unity in $\mathbb{Q}_p$ (in fact this is a
trivial consequence of Hensel's lemma \cite{Gouvea,Schikhof}; however, we
prefer to present a direct proof). There are no more roots of unity in $\mathbb{%
Q}_p.$

Let $\Gamma _p$ be the set of the $p-1$ zeroes of the polynomial $F$ where $%
F(x)=$ $x^{p-1}-1$ in $\mathbb{Q}_p.$ Then $\Gamma _p$ is closed under
multiplication and isomorphic to the multiplicative group, $\mathbb{F}_p^{*}=%
\mathbb{F}_p\backslash \{0\};$ let $\xi _i$ be the root belonging to $i+p\mathbb{Z}%
_p.$ Then $\xi _i\cdot \xi _j\in i\cdot j+p\mathbb{Z}_p$ so that $\xi _i\cdot
\xi _j=\xi _{i\cdot j\operatorname{mod}p}.$

\subsection{Continuous case-the $p-$adic power function}

We now generalize our RDS $\phi $ to the continuous case, $i.e.$ we let the
random variables $(s(\theta ^n\cdot ))_{\mathbb{Z}^{+}}$ take values in the
state space $S=\mathbb{Z}_p.$ Then we have to study properties of the $p$-adic
power function $x\mapsto x^a.$ This map is defined for $x\in 1+\mathbb{Z}_p$
and $a\in \mathbb{Z}_p$ by 
\[
x^a=\sum_{n=0}^\infty \binom an(x-1)^n, 
\]
where 
\[
\binom a0:=1,\quad \binom an:=\frac{a(a-1)...(a-n+1)}{n!},\quad n\in \mathbb{N}%
. 
\]
Here is a result which is analogous to the one in the monomial case, with
essentially the same proof as in Lemma \ref{lemma 1}.

\begin{lemma}
\label{continuous lemma}Let $x\in 1+p\mathbb{Z}_p$. Then 
\begin{equation}
\left| x^a-1\right| _p\leq \left| a\right| _p\left| x-1\right| _p,
\label{lemma}
\end{equation}
holds, with equality for $p>2$.
\end{lemma}

{\noindent \bf Proof. }
Let $a=a_0p^d$, where $a_0\in S_1(0)$, and put $\gamma=1$ and $u=x-1$.
Define $g\colon x \mapsto x^{a_0}$, and $h \colon x \mapsto x^p$. Then 
\begin{align*}
\left| g(\gamma+u)-g(\gamma)\right| _p &=\left| x^{a_0}-1\right| _p
=\left| \sum_{n=0}^\infty \binom {a_0}n(x-1)^n-1\right|
_p\\ &=\left| a_0u+o(u^2)\right| _p= \left|u\right|_p,
\end{align*}
by the strong triangle inequality. The rest of the proof is the same as in that of Lemma 
\ref{lemma 1}$\Box$

We see from the lemma that for every $x\in 1+p\mathbb{Z}_p$ the sequence $%
x^{S_n(\omega )}$ converges to $1$ if $s(\theta ^n\omega )$ belongs to $p%
\mathbb{Z}_p$ infinitely often. This is the case when $\mathbb{P}\{s(\omega )
\in p\mathbb{Z}_p)\}$ is greater than zero. To see this let us recall the
concept of recurrence. Here we follow to the classical book of P.R. Halmos \cite{Halmos}.

\vspace{1.5ex}
\noindent \textbf{Definition (Recurrent
point) }Let $(X,\mathcal{B},\mu )$ be a finite measure space. Let $B\in 
\mathcal{B}$ and let $T$ be a measure-preserving transformation. A point $x$
is said to be \emph{recurrent with respect to} $B$ if there is a natural
number $k$ for which $T^kx\in B.$

\vspace{1.5ex}
\noindent In the spirit of this definition we
have the following famous result from ergodic theory.

\begin{theorem} Recurrence Theorem. For each $B\in \mathcal{B}$ with
$\mu(B)>0$ almost every point of $B$ is recurrent with respect to $B$.
\end{theorem}

The Recurrence theorem implies a stronger version of itself. In fact, for
almost every $x$ in $B$ (with $\mu(B)>0$), there are infinitely many values of $n$ such that $%
T^nx\in B,$ see for example \cite{Halmos}. Let $B=\{\omega :s(\omega )\in p%
\mathbb{Z}_p\}$ with $\mathbb{P}(B)>0$ and let $\theta $ be the 
Markov (left) shift (which is measure-preserving). Then it follows from the recurrence
theorem that for almost every point $\omega $ in $B$ there must be an
arbitrarily large number of moments in time when the trajectory of the point 
$\omega $ is in the set $B,$ $i.e.$ for almost every $\omega \in B,$ $%
s(\theta ^n\omega )$ belongs to $p\mathbb{Z}_p$ infinitely often.
Moreover, if $\theta$ is ergodic, almost all points of the space
enter the set $B$, and of course once they are in there they will return infinitely
many times by the recurrence theorem.
In the case that $\theta$ is ergodic we then have that
$\{1\}$ is an attractor\footnote{%
See Appendix \ref{Attractor} for the definition of an attractor.}
for the RDS $\phi$ if $\mathbb{P}(s(\omega )\in p\mathbb{Z}_p)>0$.

\begin{theorem}
\label{continuous attractor}Let $\theta$ be ergodic and let
$\mathbb{P}(s(\omega )\in p\mathbb{Z}_p)>0.$ Then the set
$\{1\}$ is an attractor for the RDS $\phi$ on $X=1+p\mathbb{Z}_p.$
\end{theorem}

{\noindent \bf Proof. }We show that 
\[
\lim_{n\rightarrow \infty }\dist(\phi (n,\omega)X,\{1\})=0\quad 
\mathbb{P}-a.e. 
\]
By definition 
\begin{eqnarray*}
\dist(\phi (n,\omega )X,\{1\}) &=&\sup_{x\in 1+p\mathbb{Z}%
_p}\inf_{z\in \{1\}}\left| \phi (n,\omega )x-z\right| _p \\
&=&\sup_{x\in 1+p\mathbb{Z}_p}\left| \phi (n,\omega )x-1\right| _p
\\
&=&\sup_{x\in 1+p\mathbb{Z}_p}\left| x^{S_n(\omega )}-1\right| _p \\
&\leq &\sup_{x\in 1+p\mathbb{Z}_p}\left| S_n(\omega )\right| _p\left|
x-1\right| _p \\
&=&\left| S_n(\omega )\right| _p\frac 1p \\
&\rightarrow &0\quad \mathbb{P}-a.e.,
\end{eqnarray*}
when $n$ goes to infinity by Poincar\'{e} Recurrence Theorem, and the last
equality holds by Lemma \ref{continuous lemma}. $\Box $

\vspace{1.5ex}
\noindent Let us now return
to the discrete case.

\subsection{Attractors}

Attractors of systems like (\ref{definition of RDS}) have been studied in 
\cite{Khrennikov et al} for the case that $p$ divides at least one $s_i\in S$%
. It was shown that there are only deterministic attractors on $\mathbb{Q}_p.$
First, $\{0\}$ and the point at infinity, $\{\infty \},$ are attractors. The
points attracted to these sets are $U_{1/p}(0)=\{x\in \mathbb{Q}_p:\left|
x\right| _p\leq 1/p\}$ and $\mathbb{Q}_p\backslash \mathbb{Z}_p$ respectively. If
one of the elements in the state space $S$ of the random variable $s$ is
divisible by $p,$ then there is one more attractor on $\mathbb{Q}_p.$ This
attractor is a subset of $\Gamma _p$. In \cite{Khrennikov et al} it was
proved with the aid of Lemma \ref{lemma 1} that in the case that $p$ divides
one of the numbers in $S,$ then the points on $S_1(0)=\{x\in \mathbb{Q}%
_p:\left| x\right| _p=1\}$ are attracted to $I_s=\psi _{s_1}^{p-1}\circ ...\circ
\psi _{s_r}^{p-1}(\Gamma _p)$. The proof is based on the same procedure as in the
proof of Theorem \ref{continuous attractor}.

We now want to say something about the case when $p$ does not divide any of
the elements in the state space $S$ of the random variables.

\subsection{Random Siegel disk}

Let us introduce a generalization of Siegel disks\footnote{%
See for example \cite{Peitgen et al}.} which we call random Siegel disks. To
do this we define a metric $d$ by $d(x,A):=\inf_{a\in A}\left| x-a\right|
_p. $

\vspace{1.5ex}
\noindent\textbf{Definition (Random Siegel disk},\textbf{\ Maximal random
Siegel disk)} Let the RDS $\varphi $ be generated by a family $\psi =(\psi
_{s_1},...,\psi _{s_r})$ of monomial mappings: $\psi _{s_i}x=x^{s_i},$ in
the sense of section \ref{discrete
time
and generation}. Let $A$ be an 
\emph{invariant} set, $i.e.$ $\psi _{s_1}\circ ...\circ \psi _{s_r}(A)=A.$
Let $O$ be a subset of $\mathbb{Q}_p$ properly containing $A$. Then $O$\ is
said to be a \emph{random Siegel disk} for the RDS $\varphi $ concentrated
\emph{around} $A$ if, for almost every $\omega $, 
\[
d(x,A)=d(\varphi (n,\omega )x,A), 
\]
for every $x\in O$ and every $n\in \mathbb{Z}^{+}.$ The set $\widetilde{O}%
=\bigcup O,$ the union of all random Siegel disks around $A,$ is said to be
a \emph{maximal random Siegel disk} around $A.$ By Lemma \ref{lemma 1} we
obtain the following result.

\begin{theorem}
Let $p>2,$ see Lemma \ref{lemma 1}. Let the monomial RDS $\phi $ be generated 
by a family 
$\psi =(\psi _{s_1},...,\psi _{s_r})$ of monomial mappings where $\psi
_{s_i}x=x^{s_i}$. Let $I_s=\psi _{s_1}^{p-1}\circ ...\circ \psi_{s_r}^{p-1}(\Gamma _p)$
and suppose that $p$ does not divide any of the $s_i\in S$.
Then $\mathbb{Z}_p$ is a maximal random Siegel disk concentrated around $I_s$
for the RDS $\phi .$
\end{theorem}

{\noindent \bf Proof. }First we prove that $S_1(0)$ is a random Siegel
disk around $I_s.$ Clearly $I_s=\psi _{s_1}^{p-1}\circ ...\circ \psi _{s_r}^{p-1}
(\Gamma
_p)$ is an invariant set. Moreover, for every $x$ on the unit sphere $S_1(0)$
we have by Lemma \ref{lemma 1} for $p>2$ that 
\begin{eqnarray*}
d(x^{S_n(\omega )},I_s) &=&\inf_{a\in I_s}\left| x^{S_n(\omega )}-a\right|
_p=\inf_{a\in I_s}\left| x^{S_n(\omega )}-a^{S_n(\omega )}\right| _p \\
&=&\inf_{a\in I_s}\left| S_n(\omega )\right| _p\left| x-a\right|
_p=\inf_{a\in I_s}\left| x-a\right| _p=d(x,I_s),
\end{eqnarray*}
where the last equality holds because $p$ does not divide any of the
elements in $S$ and therefore not a product $S_n(\omega )$ so that $\left|
S_n(\omega )\right| _p=1.$ Now, $p\mathbb{Z}_p$ is also a random Siegel disk
since $x^{S_n(\omega )}\in p\mathbb{Z}_p$ for every $n$ if $x\in p\mathbb{Z}_p$
which implies that 
\[
d(x^{S_n(\omega )},I_s)=1=d(x,I_s), 
\]
for every $x\in p\mathbb{Z}_p.$ But $\mathbb{Z}_p=S_1(0)\cup p\mathbb{Z}_p$ and $%
\left| x^{S_n(\omega )}-a\right| _p\rightarrow \infty $ for every $x$
outside $\mathbb{Z}_p=B_1(0).$ Hence $\mathbb{Z}_p$ is maximal as required$.$ $%
\Box $

\section{Definition of Markovian dynamics}

Let $X=\Gamma _p.$ Given an initial state $x\in X$ $,$ our RDS defined by
the random map $\phi ,$ defined by (\ref{definition of RDS}), over a Markov
shift $\theta $ can be considered as a $\Gamma _p-$valued stochastic process
defined by the forward motion 
\begin{equation}
(x^{S_n})_{n\in \mathbb{Z}^{+}}=\left( \phi (n,\cdot )x\right) _{n\in \mathbb{Z}%
^{+}}.  \label{forward motion}
\end{equation}
We say that a sequence $(x^{S_n(\omega )})_{n=1}^N$ is an $N$ step \emph{%
realization} of the stochastic process (\ref{forward
motion})$.$ Then $%
(x^{S_n})_{n\in \mathbb{Z}^{+}}$ is a stochastic process with state space $%
\Gamma _p$ and transition probability $P(x,B)=\mathbb{P}\{\omega :x^{s(\omega
)}\in B\}$ (a proof is given in \cite{Arnold}). Thus, on $\Gamma _p$ we have
a family $(x^{S_n})_{x\in \Gamma _p}$ of stochastic processes. We want to
investigate when each process $(x^{S_n})_{n\in \mathbb{Z}%
^{+}}$ satisfies the (weak) Markov property

\begin{eqnarray}
\mathbb{P}(\phi (1+n,\omega )x &=&x_{n+1}\mid \phi (n,\omega )x=x_n,...,\phi
(1,\omega )x=x_1)  \nonumber  \label{weak Markov property} \\
&=&\mathbb{P}(\phi (1+n,\omega )x=x_{n+1}\mid \phi (n,\omega )x=x_n),
\end{eqnarray}
for every sequence $(x_i\in \Gamma _p)$ such that 
\[
\mathbb{P}(\phi (n,\omega )x=x_n,...,\phi (1,\omega )x=x_1)>0. 
\]
In doing so we define \emph{transition sets } 
\begin{equation}
A^n(x,y)=\{\alpha =\alpha _1\cdot ...\cdot \alpha _n:\alpha _i\in S\quad 
\text{and }x^\alpha =y\},  \label{transition set}
\end{equation}
of all possible ordered products of $n$ elements in $S,$ taking $x$ to $y$
in $n$ steps. With the aid of the transition sets (\ref{transition
set}) we
can write the probability of the $n$ step realization $(x_i)_{i=1}^n$ as 
\begin{eqnarray*}
\mathbb{P(}x^{S_1(\omega )} &=&x_1,\text{ }x^{S_2(\omega )}=x_2,\text{ ... },%
\text{ }x^{S_n(\omega )}=x_n) \\
&=&\mathbb{P(}x^{s(\omega )}=x_1,\text{ }x_1^{s(\theta \omega )}=x_2,\text{ ... 
},\text{ }x_{n-1}^{s(\theta ^{n-1}\omega )}=x_n) \\
&=&\mathbb{P}(s(\omega )\in A^1(x,x_1),...,\text{ }s(\theta ^{n-1}\omega )\in
A^1(x_{n-1},x_n)) \\
&=&\mathbb{P}(\omega _0\in A^1(x,x_1),...,\omega _{n-1}\in A^1(x_{n-1},x_n)).
\end{eqnarray*}
On $\Gamma _p$ the dynamics is discrete. Thus (for a sequence $(x_i)$ where $%
x_i\in \Gamma _p)$ the weak Markov property (\ref{weak
Markov
property})
is satisfied if and only if the \emph{Markov equation} 
\begin{eqnarray}
\mathbb{P}(\omega _n &\in &A^1(x_n,x_{n+1})\mid \omega _{n-1}\in
A^1(x_{n-1},x_n),\text{... ,}\omega _0\in A^1(x,x_1))  \nonumber
\label{Markov equation} \\
&=&\mathbb{P}(\omega _n\in A^1(x_n,x_{n+1})\mid \omega _0\cdot ...\cdot \omega
_{n-1}\in A^n(x,x_n)),
\end{eqnarray}
holds true for every sequence $(x_i\in \Gamma _p)$ such that 
\begin{equation}
\mathbb{P}(\omega _0\in A^1(x,x_1),\text{... ,}\omega _n\in A^1(x_{n-1},x_n))>0.
\label{required}
\end{equation}

\textbf{Remark 3} Note that on the right hand side of the Markov property
defined by (\ref{weak
Markov property}) we allow dependence on the initial
state $x.$ This is in fact the {\it weak Markov property,} see for example \cite
{Dynkin}. Another formulation of Markovian dynamics, allowing no dependence
on the past, could be: The dynamics on $X$ is Markovian under the RDS $\phi $
if 
\begin{eqnarray*}
\mathbb{P}(\phi (1+n,\omega )x &=&x_{n+1}\mid \phi (n,\omega )x=x_n,...,\phi
(1,\omega )x=x_1) \\
&=&\mathbb{P}\{\omega :\psi (\theta ^n\omega )x_n=x_{n+1}\}.
\end{eqnarray*}
In this case $\theta $ has to be a Bernoulli shift since
$\mathbb{P}\{\omega :\psi (\theta ^n\omega
)x_n=x_{n+1}\}=\mathbb{P}(\omega _n\in A^1(x_n,x_{n+1}))$ so that 
\begin{eqnarray*}
\mathbb{P}(\omega _n &\in &A^1(x_n,x_{n+1})\mid \omega _{n-1}\in
A^1(x_{n-1},x_n),\text{... ,}\omega _0\in A^1(x,x_1)) \\
&=&\mathbb{P}(\omega _n\in A^1(x_n,x_{n+1})).
\end{eqnarray*}
$\Box $

In what follows we consider Markovian dynamics in the
framework of {\it weak Markov property,} namely Markov families, see \cite{Arnold,Dynkin}. 

The family $(x^{S_n})_{x\in \Gamma _p}$ of processes is called a 
\emph{Markov family} if and only if $(x^{S_n})_{n\in \mathbb{Z}^{+}}$ 
is a Markov process for
each initial state $x\in \Gamma _p.$ We say that the \emph{dynamics} on 
$\Gamma _p$ is \emph{Markovian} if $(x^{S_n})_{x\in \Gamma _p}$ is a Markov
family\footnote{This approach is quite natural for models of the process of thinking \cite
{Khrennikov, Khrennikov et al, Albeverio Khrennikov Kloeden}. Here the
choice of the initial idea $x$ plays the crucial role.}.

We remark that the sufficient condition for Markovian dynamics is 
that $\theta $ is a Bernoulli shift. 

One may ask whether this can be
generalized directly to any Markov shift, $i.e.$ to every stochastic matrix $%
P.$ The following example illustrates that this is in fact not the case.

\noindent {\bf Example (A non-Markovian $p$-adic chain)}
Consider the RDS on $\Gamma _7.$ Let $S=\{7,2,3\}$ and let the elements of $%
S $ be distributed by $\pi =\frac 1{20}(8,9,3).$ The probability vector $\pi 
$ is a row eigenvector of the stochastic matrix 
\[
P=\left( 
\begin{array}{ccc}
\frac 12 & \frac 14 & \frac 14 \\ 
\frac 13 & \frac 23 & 0 \\ 
\frac 13 & \frac 13 & \frac 13
\end{array}
\right) . 
\]
Let $\mathbb{P}=\mu _{\pi P}$ be the corresponding Markov measure. Let $\xi $
be a primitive $6th$ root of unity in $\mathbb{Q}_7$ so that $\Gamma _7=\{1,\xi
,\xi ^2,\xi ^3,\xi ^4,\xi ^5\}.$ Note that on $\Gamma _7$ we have that $%
x^7=x^1$ for every $x$. Then consider the initial state $x_0=\xi $ and the
realization 
\[
(\xi ^3,\xi ^3,1). 
\]
Then the one step transition sets, defined by (\ref{transition set}), are $%
A^1(\xi ,\xi ^3)=\{3\},$ $A^1(\xi ^3,\xi ^3)=\{1\}$ and $A^1(\xi
^3,1)=\{2\}. $ Hence the left hand side of (\ref{Markov equation}) becomes 
\[
\mathbb{P}(\omega _2=2\mid \omega _1=1,\omega _0=3)=\frac{\mathbb{P}([3,1,2])}{%
\mathbb{P}([3,1])}=\frac{p_3p_{31}p_{12}}{p_3p_{31}}=p_{12}=\frac 14, 
\]
and the right hand side with the two step transition set $A^2(\xi ,\xi
^3)=\{3\}:$%
\begin{eqnarray*}
\mathbb{P}(\omega _2 &=&2\mid \omega _1\cdot \omega _0=3)=\frac{\mathbb{P}%
([3,1,2])+\mathbb{P}([1,3,2])}{\mathbb{P}([3,1])+\mathbb{P}([1,3])} \\
&=&\frac{p_3p_{31}p_{12}+p_1p_{13}p_{32}}{p_3p_{31}+p_1p_{13}}=\frac{8\frac
14\frac 13+3\frac 13\frac 23}{8\frac 14+3\frac 13}=\frac 49.
\end{eqnarray*}
Thus we have found a non-Markovian $p-$adic chain. $\Box $

This was the counterexample but we can ask: Is there any stochastic matrix $%
P$ which is not generating a Bernoulli shift and still satisfies the Markov
equation (\ref{Markov equation})? And, on the contrary, are there state
spaces $\Gamma _p$ and $S$ such that (\ref{Markov equation}) implies that $%
\theta $ has to be a Bernoulli shift? We will see that for some $S$ we have
to require that our Markov shift is a Bernoulli shift in order to get
Markovian dynamics.

\section{Conditions for Markovian dynamics\label{Results}}

In order to solve the Markov equation (\ref{Markov equation}) we need to
find transition sets (for possible realizations $(x_i)_{i=1}^n)$ defined by (%
\ref{transition set}). To facilitate this procedure we take advantage of the
algebraic properties of $\Gamma _p.$ It was stated in section 3 that $\Gamma
_p$ is (algebraically) isomorphic to $\mathbb{F}_p^{*},$ the multiplicative
subgroup of the residue class modulo $p.$ Hence $\Gamma _p$ is a cyclic
group under multiplication with $p-1$ elements. Thus one of the roots, $\xi
, $ is a primitive $(p-1)th$ root of unity so that $\Gamma _p=\{1,\xi
_{p-1},...,\xi _{p-1}^{p-2}\}$. Moreover every element $x\in \Gamma _p$ (in
particular the initial state of the RDS) is generating a subgroup

\[
\left\langle x\right\rangle =\{1,x,...,x^{k-1}:x^i\neq 1\text{ for }0<i<k%
\text{ and }x^k=1\}, 
\]
with $k$ elements. We say that $\left\langle x\right\rangle $ is of \emph{%
order} $k$ and let $\left| \left\langle x\right\rangle \right| $ denote the
order of $x.$ Hence an equality $x^{ab}=x^{cd}$ can be formulated as a
congruence in the sense that 
\begin{equation}
x^{ab}=x^{cd}\quad \Leftrightarrow \quad ab\equiv cd\operatorname{mod}\left|
\left\langle x\right\rangle \right| .  \label{congruence}
\end{equation}
Consequently, we can determine transition sets (\ref{transition set})
counting modulo $\left| \left\langle x\right\rangle \right| $: 
\[
A^n(x,x^\beta )=\{\alpha =\alpha _1\cdot ...\cdot \alpha _n:\alpha _i\in
S\quad \text{and }\alpha \equiv \beta \operatorname{mod}\left| \left\langle
x\right\rangle \right| \}. 
\]
\textbf{Remark 4} Given the initial state $x$ the dynamics is restricted to $%
\left\langle x\right\rangle .$ From (\ref{congruence}) it follows that the
dynamics on $\left\langle x\right\rangle $ under the RDS $\phi $ is totally
described by the dynamics on $S$ if the elements in $S\;$are considered as
elements in the residue class modulo $\left| \left\langle x\right\rangle
\right| ,$ which we denote by $\mathbb{F}_{\left| \left\langle x\right\rangle
\right| }$. Therefore the properties of the long-term behaviour of the RDS
on $\Gamma _p$ depend strongly on the order of $x.$ If the order of $x$, $%
\left| \left\langle x\right\rangle \right| ,$ is not a prime, the residue
class modulo $\left| \left\langle x\right\rangle \right| $ contains divisors
of zero, $i.e.$ there are elements $a,b\in \mathbb{F}_{\left| \left\langle
x\right\rangle \right| }$ different from $0$ in $\mathbb{F}_{\left|
\left\langle x\right\rangle \right| }$ such that $ab\equiv 0\operatorname{mod}\left|
\left\langle x\right\rangle \right| .$ Then $x^{ab}=x^0=1$ which leads to
trivial dynamics. For example $2\cdot 2=4$ which equals $0$ in $\mathbb{F}_4.$
But if $\left| \left\langle x\right\rangle \right| $ is prime then $\mathbb{F}%
_{\left| \left\langle x\right\rangle \right| }$ is a field. Thus $\mathbb{F}%
_{\left| \left\langle x\right\rangle \right| }^{*}=$ $\mathbb{F}_{\left|
\left\langle x\right\rangle \right| }\backslash \{0\}$ is a group under
multiplication and therefore contains no divisors of zero. Consequently, if $%
\left| \left\langle x\right\rangle \right| $ is a prime number and $%
S\subseteq \{1,...,\left| \left\langle x\right\rangle \right| -1\}$ the
dynamics is restricted to $\left\langle x\right\rangle \backslash \{1\}$ so
that the RDS can not enter the state $1.$ Hence, to avoid trivial dynamics
we will thus assume that the order of $x$ is a prime number
\footnote{%
The case where $\left| \left\langle x\right\rangle \right|$ is not a prime
for any $x\in \Gamma _p$ is treated in Section
\ref{when the order of x is not a prime number}.}
different from $2$ (If $\left| \left\langle
x\right\rangle \right| =2,$ $S$ will contain only one element.)
and that $S\subseteq \{1,...,\left| \left\langle
x\right\rangle \right| -1\}.$ Such an $x$ exists if $p
{\not\equiv}1\operatorname{mod}4;$ by the theorem of Lagrange we know that $\left|
\left\langle x\right\rangle \right| $ is a divisor of $\left| \Gamma
_p\right| =p-1.$ Since $\Gamma _p$ is abelian the inverse of the theorem of
Lagrange Theorem is also true, $i.e.$ given a prime divisor $n$ of $p-1$
there is a $x\in \Gamma _p$ of order $n.$ Now $p-1$ is divisible by a prime
number different from $2$ if $p{\not\equiv}1\operatorname{mod}%
4. $ $\Box $

\vspace{1.5ex}
\noindent\textbf{Remark 5} If $\left| \left\langle x\right\rangle
\right| $ is prime and $S\subseteq \{1,...,\left| \left\langle
x\right\rangle \right| -1\},$ then all one step transition sets $%
A^1(x_i,x_{i+1}),$ $x_j\in \Gamma _p$ are singletons. This is a direct
consequence of the group property of $\mathbb{F}_{\left| \left\langle
x\right\rangle \right| }^{*}.\Box $

Furthermore, it is clear that we only need to consider $n$ step conditional
probabilities in (\ref{Markov equation}) for which $n\geq 3,$ since (\ref
{Markov equation}) is always valid for $n=2.$ The following results describe
how the Markov equation (\ref{Markov equation}) is putting conditions on
the entries in the transition matrix $P.$ We shall consider the case of
Markov shifts generated by transition matrices with row eigenvectors $\pi $
where $p_i>0$ for all $i\in S.$ This is not a real restriction; since otherwise
no cylinder containing $s_i$ would have positive measure. Then the state
$s_i$ might as well be deleted. As a consequence, each row and each column
of $P$ contains a positive entry.

Also note that if the columns of $P$ are constant, then the corresponding
Markov shift is just a Bernoulli shift.

\begin{lemma}
Let $\left| \left\langle x\right\rangle \right| $ be a prime number, $%
S\subseteq \{1,...,\left| \left\langle x\right\rangle \right| -1\}$ and let $%
i_1\cdot ...\cdot i_n,$ $i_r\in S$ be an arbitrary ordered product. Then the 
$n+1$ step realization $(x^{i_1},...,x^{i_1...i_nk}),$ given the $n$ step
realization\ $(x^{i_1},...,x^{i_1...i_n})$, generates the Markov equation 
\[
p_{i_nk}=\mathbb{P}(\omega _n=k\mid \omega _0\cdot ...\cdot \omega _{n-1}\in
\{\alpha :\alpha \equiv i_1\cdot ...\cdot i_n\operatorname{%
mod}\left| \left\langle x\right\rangle \right| \}),
\]
if and only if $p_{i_1i_2}\cdot ...\cdot p_{i_{n-1}i_n}>0.$
\end{lemma}

{\noindent \bf Proof. }First we prove that (\ref{required}) is satisfied
if and only if $p_{i_1i_2}\cdot ...\cdot p_{i_{n-1}i_n}>0.$ In Remark 5 we
found that one step transition sets are singletons. Therefore 
\begin{eqnarray*}
\mathbb{P}(\omega _0 &\in &A^1(x,x^i),...,\omega _{n-1}\in A^1(x^{i_1\cdot
...\cdot i_{n-1}},x^{i_1\cdot ...\cdot i_{n-1}i_n})) \\
&=&\mathbb{P}(\omega _0=i_1,...,\omega
_{n-1}=i_n)=p_{i_1}p_{i_1i_2}\cdot ...\cdot p_{i_{n-1}i_n},
\end{eqnarray*}
which is greater than zero if and only if $p_{i_1i_2}\cdot ...\cdot
p_{i_{n-1}i_n}>0$ (we assume that $p_{i_1}>0$). For the left hand side of (%
\ref{Markov
equation}) we obtain 
\begin{align*}
\mathbb{P}(\omega _n &\in A^1(x^{i_1\cdot ...\cdot i_n},x^{i_1\cdot ...\cdot i_nk})
\mid \omega_{n-1}\in A^1(x^{i_1...i_{n-1}},x^{i_1...i_n}),\\
&\qquad\qquad\text{...,}\omega _0\in
A^1(x,x^{i_1})) \\
&=\mathbb{P}(\omega _n\in \{k\}\mid \omega _{n-1}\in \{i_n\}\text{,...,}\omega
_0\in \{i_1\}) \\
&=\mathbb{P}(\omega _n=k\mid \omega _{n-1}=i_n\text{,...,}\omega _0=i_1)=p_{i_nk}.
\end{align*}
For the right hand side we have 
\begin{equation*}
\begin{split}
\mathbb{P}(\omega _n & \in A^1(x^{i_1...i_n},x^{i_1...i_nk})\mid \omega _0\cdot
...\cdot \omega _{n-1}\in A^n(x,x^{i_1...i_n})) \\
& = \mathbb{P}(\omega _n=k\mid \omega _0\cdot ...\cdot \omega _{n-1}\in \{\alpha
: \alpha \equiv i_1\cdot ...\cdot i_n%
\operatorname{mod}\left| \left\langle x\right\rangle \right| \}),
\end{split}
\end{equation*}
as required.$\Box $

\begin{lemma}
\label{general lemma for equal rows}Let $\left| \left\langle x\right\rangle
\right| $ be a prime number, $S\subseteq \{1,...,\left| \left\langle
x\right\rangle \right| -1\}$ and let $i_1\cdot ...\cdot i_n,$ $i_r\in S$ and 
$j_1\cdot ...\cdot j_n,$ $j_r\in S$ be arbitrary ordered products. Then if $%
i_1\cdot ...\cdot i_n\equiv j_1\cdot ...\cdot j_n\operatorname{mod}\left|
\left\langle x\right\rangle \right| $ and $p_{i_1i_2}\cdot ...\cdot
p_{i_{n-1}i_n}>0$ and $p_{j_1j_2}\cdot ...\cdot p_{j_{n-1}j_n}>0$ the Markov
equation (\ref{Markov equation}) implies that 
\[
p_{i_nk}=p_{j_nk}\quad \text{for all }k\in S,
\]
$i.e.$ row $i_n$ is equal to row $j_n$ in the transition matrix $P.$
\end{lemma}

{\noindent \bf Proof. }Let $j$ be an arbitrary element in $S.$\textbf{\ }%
Consider the $n+1$ step realization $(x^{i_1},...,x^{i_1\cdot ...\cdot
i_n},x^{i_1\cdot ...\cdot i_nk}),$ given that the $n$ step realization $%
(x^{i_1},...,x^{i_1\cdot ...\cdot i_n})$ has occurred. Then from the previous
lemma we have that 
\begin{equation}
p_{i_nk}=\mathbb{P}(\omega _n=k\mid \omega _0\cdot ...\cdot \omega _{n-1}\in
\{\alpha : \alpha \equiv i_1\cdot ...\cdot i_n\operatorname{%
mod}\left| \left\langle x\right\rangle \right| \}).  \label{equation 3}
\end{equation}
Then consider the $n+1$ step realization $(x^{j_1},...,x^{j_1\cdot ...\cdot
j_n},x^{j_1\cdot ...\cdot j_nk}),$ given the realization\ $%
(x^{j_1},...,x^{j_1\cdot ...\cdot j_n}).$ According to the previous lemma 
\begin{equation}
p_{j_nk}=\mathbb{P}(\omega _n=k\mid \omega _0\cdot ...\cdot \omega
_{n-1}\in \{\alpha : \alpha \equiv
j_1\cdot ...\cdot j_n\operatorname{mod}\left| \left\langle x\right\rangle \right| \}).
\label{equation 4}
\end{equation}
Since by hypothesis $i_1\cdot ...\cdot i_n\equiv j_1\cdot ...\cdot j_n\operatorname{%
mod}\left| \left\langle x\right\rangle \right| $, the left hand side of (\ref
{equation 3}) and (\ref{equation 4}) must coincide. Consequently, $%
p_{i_nk}=p_{j_nk}$ for every $k\in S,$ as required. $\Box $

\begin{lemma}
\label{permut proposition }Let $\sigma $ be an arbitrary permutation on $%
\mathbb{F}_{2n}$ for some natural number $n.$ Then the map 
\[
\gamma _\sigma :\mathbb{F}_{2n}\rightarrow \mathbb{F}_{2n},\quad i\mapsto i+\sigma
(i),
\]
is not onto.
\end{lemma}

{\noindent \bf Proof.}\footnote{This proof is due to Robert Lagergren at the
Department of Mathematics, Statistics and Computer Science at V\"axj\"o University.}
Assume the opposite. By hypothesis 
\[
\sum_{i=0}^{2n}i\equiv \sum_{i=0}^{2n}[i+\sigma (i)]  \mod{2n}. 
\]
The left hand side of this equation is $(2n-1)n$ which is congruent to $-n$
modulo $2n.$ But the sum in the left hand side is twice this sum and thus
congruent to $0$ modulo $2n$ contrary hypothesis. $\Box $

Note that multiplication modulo $p$ is isomorphic to addition modulo $p-1$
for every prime number $p.$ Moreover $p-1$ is even, so that have

\begin{corollary}
\label{permut corollary }Let $\sigma $ be an arbitrary permutation on $\mathbb{F%
}_p^{*}.$ Then the map 
\[
\gamma _\sigma :\mathbb{F}_p^{*}\rightarrow \mathbb{F}_p^{*},\quad i\mapsto
i\sigma (i),
\]
is not onto.
\end{corollary}

\begin{theorem}
\label{2 rows are equal}Let $\left| \left\langle x\right\rangle \right| $\
be a prime number and let $S\subseteq \{1,...,\left| \left\langle
x\right\rangle \right| -1\}.$ Then at least two rows of $P$ are equal.
\end{theorem}

{\noindent \bf Proof. }As we noted before each row and each column of the
transition matrix $P$ contains a positive entry. Thus $P$ contains $\left|
\left\langle x\right\rangle \right| -1$ positive entries, $(p_{i\sigma
(i)})_{i=1}^{\left| \left\langle x\right\rangle \right| -1},$ for some
permutation $\sigma .$ Now Corollary \ref{permut corollary } implies that $%
a\sigma (a)\equiv b\sigma (b)\operatorname{mod}\left| \left\langle x\right\rangle
\right| $ for two different elements $a$ and $b$ in $\mathbb{F}_p^{*}.$ Then,
by Lemma \ref{general lemma for equal rows}, row $\sigma (a)$ equals row $%
\sigma (b).$ $\Box $

\vspace{1.5ex}
\noindent\textbf{Example }Let $\left| \left\langle
x\right\rangle \right| =5$ and $S=\{1,2,3,4\}.$ By the group property every
element in $\mathbb{F}_5^{*}$ can be written as a product of two elements in
four ways if we do care about order: 
\begin{eqnarray*}
1 &=&1\cdot 1=2\cdot 3=3\cdot 2=4\cdot 4 \\
2 &=&1\cdot 2=2\cdot 1=3\cdot 4=4\cdot 3 \\
3 &=&1\cdot 3=2\cdot 4=3\cdot 1=4\cdot 2 \\
4 &=&1\cdot 4=2\cdot 2=3\cdot 3=4\cdot 1.
\end{eqnarray*}
Then if $p_{11},p_{23},p_{32},p_{44}>0$ we have according to Lemma \ref
{general lemma for equal rows} that $p_{1k}=p_{3k}=p_{2k}=p_{4k}$ so that
the rows of the transition matrix $P\;$ are constant. Thus we obtain the
following result.

\begin{theorem}
Let $\left| \left\langle x\right\rangle \right| $\ be a prime number and let 
$S\subseteq \{1,...,\left| \left\langle x\right\rangle \right| -1\}.$\ If $%
\left| \left\langle x\right\rangle \right| -1$\ entries of the transition
matrix $P$\ are greater than zero and the product of the indeces for each of
these entries are equal, then $(x^{S_n})_{x\in X}$ is a Markov family if and
only if $\theta $ is a Bernoulli shift.
\end{theorem}

Let us now study the case when $S=\{a,b\}\subset \{1,...,\left| \left\langle
x\right\rangle \right| -1\}$ (and $\left| \left\langle x\right\rangle
\right| $ is a prime). Then we define the transition matrix $P,$ 
\[
P=\left( 
\begin{array}{cc}
p_{aa} & p_{ab} \\ 
p_{ba} & p_{bb}
\end{array}
\right) , 
\]
generating the Markov measure $\mathbb{P}=\mu _{\pi P}.$ We obtain the
following result

\begin{theorem}
Let $\left| \left\langle x\right\rangle \right| $ be a prime number and let $%
S=\{a,b\}$ be a subset of $\{1,...,\left| \left\langle x\right\rangle \right| -1\}.$
Then $(x^{S_n})$ is a Markov process if and only if $\theta $ is a Bernoulli
shift.
\end{theorem}

{\noindent \bf Proof. }As stated before, it is necessary that each row and
each column contains a positive entry. By Lemma \ref{general lemma for equal rows}
it is clear that if $p_{ab},p_{ba}>0$ then $p_{bk}=p_{ak}$ so
that $\theta $ has to be a Bernoulli shift. In the remaining cases we must
have $p_{aa},p_{bb}>0.$ But since $\left| \left\langle x\right\rangle
\right| $ is a prime number and $a,b\neq 0$ we have (by the little theorem
of Fermat) that $a^{\left| \left\langle x\right\rangle \right| -1}\equiv
b^{\left| \left\langle x\right\rangle \right| -1}\equiv 1\operatorname{mod}\left|
\left\langle x\right\rangle \right| .$ Hence by Lemma \ref{general lemma for
equal rows} we have $p_{ak}=p_{bk}$ so that $\theta $ is a Bernoulli shift
as required. $\Box $

\subsection{The general case for $2\times 2$ matrices}

We now go on to study the general case when the order of $x$ need not be an odd prime. 
The case when the order of $x$ is not a prime is in general 
more difficult since the mappings $\psi_{s_j}:$  $x\mapsto x^{s_j}$ do not form 
a group in this case. 
We can, however, obtain some results for $2\times 2$ matrices, in other words, 
when the RDS is generated by two maps.  

Let $S=\{a,b\}$ for two natural numbers $a$ and $b$ such that they are
distinct when considered as elements in $\mathbb{F}_{\left| \left\langle
x\right\rangle \right| }$. For simplicity we first study the case when $b=p.$
First we observe that for $p>2$ we have $p^n\equiv 1\operatorname{mod}p-1$ for every
natural number $n.$ Moreover $a^mp^n\equiv a^m\operatorname{mod}p-1,$ and in fact,
since $\left| \left\langle x\right\rangle \right| $ is a divisor of $p-1:$

\begin{description}
\item  $(i)$\quad $p^n\equiv 1\operatorname{mod}\left| \left\langle x\right\rangle
\right| ,$

\item  $(ii)$ $a^mp^n\equiv a^m\operatorname{mod}\left| \left\langle x\right\rangle
\right| .$
\end{description}

Both $(i)$ and $(ii)$ are direct consequences of the rule: $x\equiv y\operatorname{%
mod}n$ implies $cx\equiv cy\operatorname{mod}n$ for any integer $c$. Let us do the
following remarks.

\vspace{1.5ex}
\noindent\textbf{Remark 6} If $p=2,$ $\Gamma _p$ contains only
one element, $1.$ Therefore every Markov shift will do. The dynamics is also
trivial for $\left| \left\langle x\right\rangle \right| =2.$ Therefore we
shall always assume that $\left| \left\langle x\right\rangle \right|
\geqslant 3.$ For $a\equiv p\equiv 1\operatorname{mod}\left| \left\langle
x\right\rangle \right| $ the dynamics is also trivial, $x^{S_n(\omega )}=x,$
so that $\phi $ will be the identity map on $\Gamma _p$. Consequently,
condition (\ref{Markov equation}) is valid (with probabilities which are equal to $%
1)$ for every possible $n$ step realization. Therefore any Markov shift will
imply that $(x^{S_n})$ is a Markov process. Also for $a$ satisfying $%
a^2\equiv a\operatorname{mod}\left| \left\langle x\right\rangle \right| $ (implying $%
a^n\equiv a\operatorname{mod}\left| \left\langle x\right\rangle \right| $ for $%
\forall n\in \mathbb{N)}$ the sequences $(x^{S_n})$ are Markov chains. $\Box $

Let us consider the case when $a{\not\equiv}1\operatorname{mod}%
\left| \left\langle x\right\rangle \right| $ and $a^2
{\not\equiv}a\operatorname{mod}\left| \left\langle x\right\rangle \right| .$ Then we
obtain the following result.

\begin{lemma}
\label{pba>0}Let $S=\{a,p\}$ where 
\begin{equation}
\left\{ 
\begin{array}{cc}
a{\not\equiv}1\operatorname{mod}\left| \left\langle
x\right\rangle \right| , &  \\ 
a^2{\not\equiv}a\operatorname{mod}\left| \left\langle
x\right\rangle \right| , & 
\end{array}
\right. .  \label{condition lemma 1}
\end{equation}
and let $p_{ap},p_{pa}>0.$ Then $(x^{S_n})$ is a Markov process if and only
if $\theta $ is a Bernoulli shift.
\end{lemma}

{\noindent \bf Proof. }Suppose that $(x^{S_n})$ is $a$ Markov process.
First note that we assume that $p_a,p_b>0.$ Let $p_{pa}>0$ and consider the
realization 
\[
(x,x^a,x^{a^2}). 
\]
Then, by the condition (\ref{condition lemma 1}), we obtain transition sets $A^1(x,x)=\{p\}$,
$A^1(x,x^a)=\{a\}$ and $A^1(x^a,x^{a^2})=\{a\}.$ Hence the
left hand side in the Markov condition (\ref{Markov equation}) is: 
\[
\Delta _1^3=\mathbb{P}(\omega _2=a\mid \omega _1=a,\omega _0=p)=\frac{\mathbb{P}%
([p,a,a])}{\mathbb{P}([p,a])}=\frac{p_pp_{pa}p_{aa}}{p_pp_{pa}}=p_{aa}. 
\]
For the right hand side of the Markov condition we use $A^2(x,x^a)=\{p\cdot
a,a\cdot p\}$ and obtain 
\begin{eqnarray*}
\Delta _2^3 &=&\mathbb{P}(\omega _2=a\mid \omega _1\cdot \omega _0=a\cdot p) \\
&=&\frac{\mathbb{P(}[p,a,a])+\mathbb{P(}[a,p,a])}{\mathbb{P}([p,a])+\mathbb{P(}[a,p])}=%
\frac{p_pp_{pa}p_{aa}+p_ap_{ap}p_{pa}}{p_pp_{pa}+p_ap_{ap}}.
\end{eqnarray*}
Now, the Markov condition $\Delta _1^3=\Delta _2^3$ implies that 
\[
p_ap_{ap}p_{pa}=p_ap_{ap}p_{aa}. 
\]
By the condition of the lemma we conclude that $p_{aa}=p_{pa}.$ Hence, the
columns of $P$ are constant and $\theta $ is a Bernoulli shift. $%
\Box $

Note that $p_{pa}=0$ implies that $p_{pp}=1$ so that $p$ is an absorbing
state. In this case $P$ is reducible. Also note that if $p_{ap}=0$ the last
equality in the proof does not give any condition. Now, by the previous
Lemma we obtain the following result.

\begin{theorem}
\label{p theorem}Let $S=\{a,p\}$ where 
\begin{equation}
\left\{ 
\begin{array}{cc}
a{\not\equiv}1\operatorname{mod}\left| \left\langle
x\right\rangle \right| , &  \\ 
a^2{\not\equiv}a\operatorname{mod}\left| \left\langle
x\right\rangle \right| , &  \\ 
a^n\equiv 1\operatorname{mod}\left| \left\langle x\right\rangle \right| , & \text{%
for some }n\geqslant 2.
\end{array}
\right.   \label{2x2 always Bernoulli case}
\end{equation}
Then $(x^{S_n})$ is a Markov process if and only if $\theta $ is a Bernoulli
shift.
\end{theorem}

{\noindent \bf Proof. }Suppose that $(x^{S_n})$ is a Markov process. By
the previous lemma we can assume that $p_{ap}$ or $p_{pa}$ equals zero. Now
let $m=\min \{n:$ $a^n\equiv 1\operatorname{mod}\left| \left\langle x\right\rangle
\right| \}.$ Consider the following three cases:

\begin{description}
\item  1) Let $p_{ap}>0$ and $p_{pa}=0,$ so that $p_{pp}=1.$ Consider the $%
m+1$ step realization 
\[
(x,...,x,x^a).
\]
Then by the condition of the Theorem we have transition sets $%
A^1(x_i,x_{i+1})=\{p\}$ for $0\leq $ $i\leq m$ and $A^1(x_m,x_{m+1})=\{a\}$.
Therefore the left hand side of (\ref{Markov equation}) is 
\[
\Delta _1^{m+1}=\frac{\mathbb{P(}[p,...,p,a])}{\mathbb{P(}[p,...,p])}=\frac{%
p_pp_{pp}...p_{pp}p_{pa}}{p_pp_{pp}...p_{pp}}=p_{pa},
\]
and since $A^m(x,x)=\{a^m,p^m\}$ the right hand side of (\ref
{Markov
equation}) becomes 
\begin{align*}
  \Delta _2^{m+1} & =\frac{\mathbb{P(}[p,...,p,a])+\mathbb{P(}[a,...a,a])}{\mathbb{P(}%
  [p,...,p])+\mathbb{P(}[a,...,a])}\\
  & =\frac{p_pp_{pp}...p_{pp}p_{pa}+p_ap_{aa}...p_{aa}p_{aa}}{%
  p_pp_{pp}...p_{pp}+p_ap_{aa}...p_{aa}}.
\end{align*}
Now the Markov condition $\Delta _1^{m+1}=\Delta _2^{m+1}$ implies that 
\begin{equation}
p_ap_{aa}...p_{aa}p_{aa}=p_ap_{aa}...p_{aa}p_{pa},  \label{villkor}
\end{equation}
so that $p_{aa}=0$ (since $p_{pa}=0).$ Thus the columns of $P$ are constant
and consequently $\theta $ is a Bernoulli shift .

\item  2) Let $p_{ap}=p_{pa}=0.$ Then we can consider the above given
realization. Consequently (\ref{villkor}) is not valid since $p_{aa}>0$
implies that the left hand side of (\ref{villkor}) is positive while $%
p_{pa}=0$ implies that the right hand side is zero.

\item  3) Let $p_{ap}=0$ and $p_{pa}>0,$ so that $p_{aa}=1.$ Consider the $%
m+1$ step realization 
\begin{equation}
(x^a,x^{a^2}...,x^{a^{m-1}},x,x).  \label{p theorem realization}
\end{equation}
Then by the condition of the Theorem we have transition sets $%
A^1(x_i,x_{i+1})=\{a\}$ for $0\leq $ $i\leq m$ and $A^1(x_m,x_{m+1})=\{p\}$.
Therefore the left hand side of (\ref{Markov
equation}) is 
\[
\Delta _1^{m+1}=\frac{\mathbb{P(}[a,...,a,p])}{\mathbb{P(}[a,...,a])}=\frac{%
p_ap_{aa}...p_{aa}p_{ap}}{p_ap_{aa}...p_{aa}}=p_{ap}, 
\]
and since $A^m(x,x)=\{a^m,p^m\}$ the right hand side of (\ref
{Markov
equation}) becomes 
\begin{align*}
  \Delta _2^{m+1} & = \frac{\mathbb{P}([a,...a,p])+\mathbb{P}([p,...,p,p])}{\mathbb{P}(%
  [a,...,a])+\mathbb{P(}[p,...,p])}\\
  & =\frac{p_ap_{aa}...p_{aa}p_{ap}+p_pp_{pp}...p_{pp}p_{pp}}{%
  p_ap_{aa}...p_{aa}+p_pp_{pp}...p_{pp}}. 
\end{align*}
Now the Markov condition $\Delta _1^{m+1}=\Delta _2^{m+1}$ implies that 
\[
p_pp_{pp}...p_{pp}p_{pp}=p_pp_{pp}...p_{pp}p_{ap,} 
\]
so that $p_{pp}=p_{ap}$ (since $p_{pp}>0).$ Thus the columns of $P$ are
constant in every case and consequently $\theta $ is a Bernoulli shift . $%
\Box $
\end{description}

Note that $a\equiv -1\operatorname{mod}\left| \left\langle x\right\rangle \right| $
is a solution to (\ref{2x2 always Bernoulli case}) for $\left| \left\langle
x\right\rangle \right| \geqslant 3.$ Thus we have:

\begin{corollary}
Let $S=\{a,p\}.$ Then if $a$ $\equiv -1\operatorname{mod}\left| \left\langle
x\right\rangle \right| $ fore some $x\in \Gamma _p$ with $\left|
\left\langle x\right\rangle \right| \geqslant 3,$ the dynamics on $\Gamma _p$
is Markovian if and only if $\theta $ is a Bernoulli shift.
\end{corollary}

We now study the dynamics on $\Gamma _p$ when $S=\{a,b\}$ and $p$ is a
divisor of $b.$ This is in fact equivalent to the case when $b\in \mathbb{N}$
is arbitrary, since $b=kp,$ $k\in \mathbb{N}$ implies that $b^n\equiv
k^np^n\equiv k^n\operatorname{mod}\left| \left\langle x_0\right\rangle \right| $ $%
\forall n\in \mathbb{N}.$ Hence we study $S=\{a,b\},$ $a,b\in \mathbb{N}.$

\vspace{1.5ex}
\noindent\textbf{Remark 7} Our previous results for $b=p$ can be generalized directly
to the case when $b\equiv 1\operatorname{mod}\left| \left\langle x\right\rangle
\right| .$ $\Box $

For $b=p$ we found that if $a$ satisfies (\ref{condition lemma 1}) and $%
p_{ap},p_{pa}>0,$ then the Markov shift $\theta $ has to be a Bernoulli
shift. This result can be generalized according to:

\begin{lemma}
Let $S=\{a,b\}$ where 
\begin{equation}
\left\{ 
\begin{array}{cc}
a{\not\equiv}b\operatorname{mod}\left| \left\langle
x\right\rangle \right| , &  \\ 
ab{\not\equiv}b^2\operatorname{mod}\left| \left\langle
x\right\rangle \right| , &  \\ 
a^2{\not\equiv}ab\operatorname{mod}\left| \left\langle
x\right\rangle \right| , &  \\ 
a^2b{\not\equiv}ab^2\operatorname{mod}\left| \left\langle
x\right\rangle \right| . & 
\end{array}
\right. .  \label{pba>0 condition}
\end{equation}
Then if $p_{ab},p_{ba}>0,$ $(x^{S_n})$ is a Markov process if and only if $%
\theta $ is a Bernoulli shift.
\end{lemma}

{\noindent \bf Proof. }Replacing $p$ by $b,$ and considering the
realization $(x^b,x^{ba},x^{ba^2})$ the proof is identical to that of Lemma 
\ref{pba>0}$.$ $\Box $

It is clear that $a\equiv 2\operatorname{mod}\left| \left\langle x\right\rangle
\right| $ and $b\equiv 1\operatorname{mod}\left| \left\langle x\right\rangle \right| 
$ are solutions of (\ref{pba>0 condition}). The same holds for $a\equiv -1%
\operatorname{mod}\left| \left\langle x\right\rangle \right| .$ In Appendix \ref
{solutions} we have shown that the numbers $a\equiv -2\operatorname{mod}\left|
\left\langle x\right\rangle \right| $ and $b\equiv -1\operatorname{mod}\left|
\left\langle x\right\rangle \right| $ also satisfy (\ref{pba>0 condition}).

We may ask for the existence of more solutions. The answer is that these are
the only general solutions for $\left| \left\langle x_0\right\rangle \right|
\geqslant 3.$ But of course the number of solutions may far exceed the
one given above
even for small orders of $\left\langle x\right\rangle .$ For $\left|
\left\langle x\right\rangle \right| =5$ we have that every combination of $a$
and $b$ for which $a{\not\equiv}b\operatorname{mod}\left|
\left\langle x\right\rangle \right| ,$ satisfies the condition (\ref
{pba>0
condition}). Whereas for $\left| \left\langle x\right\rangle \right|
=6$ there only exists one more solution, $a\equiv 2\operatorname{mod}\left|
\left\langle x\right\rangle \right| $ and $b\equiv 4\operatorname{mod}\left|
\left\langle x\right\rangle \right| ,$ which is, however, not a solution for 
$\left| \left\langle x\right\rangle \right| =8.$

We now give a generalization of Theorem \ref{p theorem}.

\begin{theorem}
\label{general 2x2 theorem}Let $S=\{a,b\}$ and suppose $(a,b)$ satisfies (%
\ref{pba>0 condition}) with the additional condition 
\[
a^n\equiv b^n\operatorname{mod}\left| \left\langle x\right\rangle \right| ,\quad
a^{n+1}{\not\equiv}b^{n+1}\operatorname{mod}\left| \left\langle
x\right\rangle \right| ,
\]
for some $n\geqslant 2.$ Then $(x^{S_n})$ is a Markov process if and only if 
$\theta $ is a Bernoulli shift.
\end{theorem}

{\noindent \bf Proof. }We replace $p$ by $b$ and replace the condition $%
a^n\equiv b^n\operatorname{mod}\left| \left\langle x\right\rangle \right| $ by $%
a^n\equiv b^n\operatorname{mod}\left| \left\langle x\right\rangle \right| $ in
Theorem \ref{p theorem}. Then, if we consider the realizations $%
(x^b,...,x^{b^{m-1}},x^{a^m},x^{a^{m+1}})$ and $%
(x^a,...,x^{a^{m-1}},x^{b^m},x^{b^{m+1}})$ respectively (for the case 1) and
3) in the proof of Theorem \ref{p theorem} ), the proof can be completed
with the same procedure as in the proof of Theorem \ref{p theorem}. $\Box $

\section{Concluding remarks}

Given a transition matrix $P\;$and a prime $p,$ $p
{\not\equiv}1\operatorname{mod}4,$ Lemma \ref{general lemma for equal rows} was
found as a useful tool for deciding whether the dynamics on $\Gamma _p$ is
Markovian or not. But this result is not just about RDS (defined on $\Gamma
_p)$ generated by monomial mappings.

Let $\left| \left\langle x\right\rangle \right| $ be a prime number and let $%
S=\{1,...,\left| \left\langle x\right\rangle \right| -1\}$. To each $a\in S$
there is a corresponding monomial mapping $\psi _a:x\mapsto x^a$ and vice
versa. Moreover by (\ref{congruence}) we have 
\[
\psi _a\circ \psi _bx=\psi _c\circ \psi _dx\quad \Leftrightarrow \quad
ab\equiv cd\operatorname{mod}\left| \left\langle x\right\rangle \right| . 
\]
Hence the composition of mappings in $(\psi _s)_{s\in S}$ is a binary
operation with the same properties as multiplication in $\mathbb{F}_{\left|
\left\langle x\right\rangle \right| }^{*}.$ Consequently, the map 
\[
\gamma :\mathbb{F}_{\left| \left\langle x\right\rangle \right| }^{*}\rightarrow
(\psi _s)_{s\in S},\quad s\mapsto \psi _s, 
\]
is an (algebraic) isomorphism. In this way the $(\psi _s)_{s\in S}$ form a
group of mappings on $\left\langle x\right\rangle \backslash \{1\}\footnote{%
Remember that the dynamics is restricted to $\left\langle x\right\rangle
\backslash \{1\}$ when $S=\{1,...,\left| \left\langle x\right\rangle \right|
-1\}$ and $\left| \left\langle x\right\rangle \right| $ is prime, see Remark
3.}$ isomorphic to $\mathbb{F}_{\left| \left\langle x\right\rangle \right|
}^{*}.$ Note that in this case the family $(\psi _s)_{s\in S}$ is in fact a
subgroup of $perm(\left\langle x\right\rangle \backslash \{1\})$, the group
of all permutations on $\left\langle x\right\rangle \backslash \{1\}$ (or on 
$\left| \left\langle x\right\rangle \right| -1$ letters). Therefore we can
consider the RDS $\phi $ on $\left\langle x\right\rangle \backslash \{1\}$
as generated by a group of permutations on $\left\langle x\right\rangle
\backslash \{1\}$ (or on $\left| \left\langle x\right\rangle \right| -1$
letters).

We are now in a position to make some more general statements. The idea is
the following. Let $X$ be a finite state space and let the family $\psi
=(\psi _s)_{s\in S}$ of mappings be a subgroup of $perm(X)\footnote{%
These are automatically measurable since in the finite case $X$ is endowed
with the $\sigma -$algebra consisting of all subsets of $X.$}$ isomorphic
to $\mathbb{F}_p^*$ and let $%
\theta $ be a Markov shift on $S^{\mathbb{N}}.$ Then consider the RDS $\varphi $
generated by $\psi $ (in the sense of section \ref
{discrete
time
and
generation}). Define transition sets $%
A^n(x,y)=\{i_1\cdot ...\cdot i_n:\psi _{i_n}\circ ...\circ \psi
_{i_1}x=y,\quad i_k\in S\}.$ Then a corresponding stochastic process $%
(\varphi (n,\cdot )x)_{n\in \mathbb{Z}^{+}}$ is a Markov process if and only if
the Markov equation (\ref{Markov equation}) holds true. Now, with just a
slight modification (not counting modulo $\left| \left\langle x\right\rangle
\right| $ but operating with the binary operation of composition on $perm(X)$%
), we obtain a result analogous to the one in Lemma \ref{general lemma
for equal rows}.

\vspace{1.5ex}
\noindent\textbf{Lemma 5.2'} \textit{Let }$%
i_1,...,i_n,$\textit{\ }$i_r\in S$\textit{\ and }$j_1,...,j_n,$\textit{\ }$%
j_r\in S$\textit{\ be arbitrary. Then if }$\psi _{i_n}\circ ...\circ \psi
_{i_1}x=\psi _{j_n}\circ ...\circ \psi _{j_1}x$\textit{\ and }$%
p_{i_1i_2}\cdot ...\cdot p_{i_{n-1}i_n}>0$\textit{\ and }$p_{j_1j_2}\cdot
...\cdot p_{j_{n-1}j_n}>0$\textit{\ the Markov equation (\ref
{Markov
equation}) implies that } 
\[
p_{i_nk}=p_{j_nk}\quad \text{for all }k\in S, 
\]
$i.e.$\textit{\ row }$i_n$\textit{\ is equal to row }$j_n$\textit{\ in the
transition matrix }$P.$
\vspace{1.5ex}
\\In section \ref{Results}, Theorem \ref{2 rows are equal},
we found that, requiring Markovian dynamics, at least
two rows of $P$ had to be equal. In fact we propose a much stronger version
of this theorem.

\vspace{1.5ex}
\noindent\textbf{Theorem 5.1'} \textit{Let }$\left| \left\langle
x\right\rangle \right| $\textit{\ be a prime number and let }$S\subseteq
\{1,...,\left| \left\langle x\right\rangle \right| -1\}.$\textit{\ Then }$%
(x^{S_n})$\textit{\ is a Markov family if and only if all rows of }$P$%
\textit{\ are equal, }$i.e.$\textit{\ }$\theta $\textit{\ is a Bernoulli
shift.}

\newpage
\appendix

\section{Attractors}\label{Attractor} 

Here we give the definition of an attractor via convergens in probability
which is worked out in the paper, \cite{Ochs}, of Ochs.
Let $\varphi\colon\mathbb{T}\times\Omega\times X\rightarrow X$ be a RDS
on the metric space $X$ with metric $d$. A \emph{random set} is a set 
$B\in \mathcal{F}\otimes\mathcal{B}$ such that $\omega\mapsto d(x,B(\omega))$
is measurable for every $x\in X$, where $B(\omega)$ denotes the \emph{section}
$B(\omega):=\{x\in X\colon (\omega,x)\in B)\}$ and 
$d(x,B(\omega)):=\inf\{d(x,y)\colon y\in B(\omega)\}$. $B$ is said to be a
\emph{random compact set}, if each $B(\omega)$ is a compact subset of $X$.
A set $B\subset\Omega\times X$ is a random compact set if and only if the 
$B(\omega)$ are compact and $\omega\mapsto B(\omega)$ is measurable with respect
to the Borel $\sigma-$algebra generated by the Hausdorff distance between compact 
subsets of $X$. This metric we denote by $\dist$ so that
$\dist(A,B):=\sup_{x\in A}d(x,B)$.\\
\textbf{Definition (\textbf{Attractor},\textbf{\ Basin of attraction})}
Let $B\subset\Omega\times X$ be a random set. A random compact set $A\subset B$
is called a \emph{weak random $B$ attractor}, if 
\begin{enumerate}[1)]
   \item $A$ is strictly forward invariant, 
   $i.e.$ $\varphi(t,\omega)A(\omega)=A(\theta(t)\omega)$ for
   every $\omega\in\Omega$ and $t>0$, 
   \item $A$ attracts random compact sets in probability, $i.e.$
\begin{align*}
\lim_{t\rightarrow\infty}\mathbb{P}\{\omega &\colon\dist(\varphi(t,\omega)C(\omega)
,A(\theta(t)\omega)) > \epsilon \}\\ 
&= \lim_{t\rightarrow\infty}\mathbb{P}\{\dist(\theta(t)C,A)> \epsilon)=0,
\end{align*}
for every random compact set $C\subset B$ and every $\epsilon>0$.
\end{enumerate}
A maximal set $B$
with the property of $A$ being a $B$ attractor is called the 
\emph{basin of attraction} of $A$.

\section{Solutions of congruences\label{solutions}}

We discuss the solutions to the system (\ref{pba>0 condition}) of
congruences for $\left| \left\langle x_0\right\rangle \right| \geqslant 3.$
It is clear that $a\equiv 2\operatorname{mod}\left| \left\langle x\right\rangle
\right| $ and $b\equiv 1\operatorname{mod}\left| \left\langle x\right\rangle \right| 
$ are solutions of (\ref{pba>0 condition}). The same holds for $a\equiv -1%
\operatorname{mod}\left| \left\langle x\right\rangle \right| .$

\begin{proposition}
The numbers $a\equiv -2\operatorname{mod}\left| \left\langle x\right\rangle \right| $
and $b\equiv -1\operatorname{mod}\left| \left\langle x\right\rangle \right| $ are
solutions to (\ref{pba>0 condition}) (for $\left| \left\langle
x\right\rangle \right| \geqslant 3).$
\end{proposition}

{\noindent \bf Proof. }It is clear that the first condition is satisfied.
The other two conditions are valid according to the considerations below.

\begin{description}
\item  (ii) $ab\equiv (-2)(-1)\equiv 2\operatorname{mod}\left| \left\langle
x\right\rangle \right| $

$b^2\equiv (-1)^2\equiv 1\operatorname{mod}\left| \left\langle x\right\rangle
\right| $

\item  (iii) $a^2b\equiv (-2)^2(-1)\equiv -4\operatorname{mod}\left| \left\langle
x\right\rangle \right| $

$ab^2\equiv (-2)(-1)^2\equiv -2\equiv \left| \left\langle x\right\rangle
\right| -2\operatorname{mod}\left| \left\langle x\right\rangle \right| $
\  $\Box $
\end{description}

\end{document}